# Concurrent *operando* neutron imaging and diffraction analysis revealing spatial lithiation phase evolution in an ultra-thick graphite electrode

Markus Strobl[1,2*], Monica E. Baur[1,3], Stavros Samothraktis[1], Florencia Malamud[1], Xiaolong Zhang[1], Patrick K.M. Tung[4], Søren Schmidt[5], R. Woracek[6], J. Lee[1], Ryoji Kiyanagi[7], Luise Theil Kuhn[3], Inbal Gavish Segev[8,9], Yair Ein-Eli[9,10,11]

[1]PSI Center for Neutron and Muon Sciences, 5232 Villigen, PSI, Switzerland
[2]Niels Bohr Institut, University of Copenhagen, Denmark
[3] Department of Energy Conversion and Storage, Technical University of Denmark, DK-2800 Kongens Lyngby, Denmark
[4]Research Technology Services (ResTech), UNSW Sydney, Kensington 2052, Australia
[5] DMSC, European Spallation Source, Lyngby, Denmark
[6]European Spallation Source, P.O. Box 176, S - 221 00 Lund, Sweden
[7] J-PARC Center, Japan Atomic Energy Agency, Tokai, 319-1195, Japan
[8]Soreq Nuclear Research Center (SNRC), Yavne, Israel
[9]Department of Material Science and Engineering, Technion–Israel Institute of Technology, 3200003 Haifa, Israel
[10]Grand Technion Energy Program (GTEP), Technion 3200003 Haifa, Israel
[11]Israel National Institute for Energy Storage (INIES), Technion 3200003 Haifa, Israel

**Abstract**

Energy efficient, safe and reliable Li-ion batteries (LIBs) are required for a wide range of applications. Charging capabilities of thick electrodes still holding their stored high-energy is a most desirable characteristic in future advanced LIBs. The introduction of ultra-thick graphite anode meets limitations in internal electrode transport properties, leading to Li-ion gradients with detrimental consequences for battery cell performance and lifetime. Yet, there is a lack of experimental tools capable of providing a complete view of local processes and evolving gradients within such thick electrodes. Here, we introduce a multi-modal *operando* measurement approach, enabling quantitative spatio-temporal observations of Li concentrations and intercalation phases in ultra-thick, graphite electrodes. Neutron imaging and diffraction concurrently provide correlated information from the macroscopic scale of the cell and electrode down to the crystallographic scale portraying the intercalation and deintercalation processes. In particular, the evolving formation of the solid electrolyte interphase (SEI), observation of gradients in total lithium content, as well as in the formation of ordered $Li_xC_6$ phases and trapped lithium have been mapped throughout the first charge-discharge cycle of the cell. Different lithiation stages co-exist during charging and discharging of an ultra-thick composite graphite-based electrode; delayed lithiation and delithiation processes are observed at the central region of the electrode, while the SEI formation, potential plating and dead lithium are predominantly found closer to the interface with the separator. The study furthermore emphasizes the potential of the method to study Li ion diffusion and the kinetics of lithiation phase formation in advanced ultra-thick electrodes. Specifically, the relation of dead Li and SEI formation to the different Li phase stages can be mapped.

## 1. Introduction

Demands on battery performance are ever-growing and requirements of electric vehicles, portable electronics as well as stationary energy storage units dictate major technological advances. Li-ion batteries (LIBs) are dominating the rechargeable battery market due to their favorable energy density, and long cycle life [1-3]. While electrode materials and battery chemistry are designed towards achieving higher energy densities of LIB [4-6], another approach is the exploration of advanced battery configurations with increased electrochemically active material ratio *via* electrode architecture engineering. However, on the one hand reduction in thickness of other, inactive structural components such Al, Cu metal current collectors, and separators and on the other hand, the two active battery materials meet technical, safety and performance related limitations. Therefore thick 3D electrode designs (around 200 µm) are being researched for LIBs production [7,8]. This approach appears to be promising, leading to an increase in the overall active electrode material portion, thus paving a way to substantially improving the overall energy density on the cell level [9-14]. It is also anticipated that this trend will continue, and one may expect further increase in the thickness of the electrodes, even reaching thicknesses of 300-400 µm in a few years. However, increasing electrode thickness may introduce transport limitations; hence, one may expect a growing cell resistance, and a substantially reduced power capability, manifested in a lower C-rate capability [15,16]. Due to extended transport paths through a thick electrode, both charge and mass transport can be severely impacted already at low C-rates, limiting the electrode operation. Even more than for thin electrodes, the capacity, thermal stability, and cycle life of the high energy LIBs relying on a thick electrode configuration, degrade at increasing cycling rates due to kinetic losses, significant $Li^+$ concentration gradients, leading to cell polarization and possibly even to a lithium plating [15,17,18]. This triggers endeavors into sophisticated material designs to improve the performance and stability of thick high energy electrodes [10-14,19-23]. The understanding of the kinetics of the processes leading to these limitations is crucial to the design of improved thick electrodes.

While the electrochemical performance of different cell configurations and chemistries can be assessed through experiments straightforwardly, the actual understanding at the microscopic level heavily relies on modelling with limited means of direct process observations, mainly due to characterization-related challenges [24]. The key processes of Li-ion transport and intercalation are generally investigated and elucidated based on theoretical studies and simulations, according to Newman-type models [25,26]. However, theoretical investigations also convey basic performance and equivalent circuits, as well as thermal and electrochemical models [27-32]. These enabled the processes of Li dendrite formation [33], Li plating [34,35], and Solid Electrolyte Interphase (SEI [36]) growth [37,38] to be simulated in detail and, thus, a better understanding of performance and lifetime limitations of cells can be envisioned. However, the lack of experimental access to corresponding details in operating cells limits full understanding and, thus, coherent modelling of these processes.

Focusing on the spatially and temporally resolved Li-ion transport and intercalation enables one to elucidate the formation of Li-ion concentration gradients and the evolution of lithiation phases and their distributions, as being intricate functions of electrode thickness ($t$), tortuosity ($\tau$), and porosity ($\varepsilon$). In addition, one can also deduce intercalation rates, ion transference numbers and Li-ion diffusion coefficients [15,17,25]. While Li-ion transport, lithiation gradients, and SEI formation are studied extensively through simulation, means for direct operando observation of these molecular multiphase and multiscale transport processes are still limited. X-ray and neutron diffraction provide information on the development of lithiation phases but generally lack the resolution to observe lithium transport and lithiation process and the evolution of gradients across an electrode. However, neutron imaging and scanning synchrotron X-ray diffraction have recently been reported to enable such observations to some extent [39-46]. However, x-rays only probe crystallographic phase fractions rather than Li concentration per se, while conventional neutron imaging observes Li distributions but not specifically intercalation and lithiation phases in the electrode.

Here, we report on our efforts to overcome these limitations via concurrent operando neutron imaging and diffraction. We utilize a time-of-flight (ToF) diffractometer at a pulsed neutron source (J-PARC, Japan) and retrofit it with a near-field transmission imaging detector to probe the SEI formation and lithiation/deliathiation processes of an ultra-thick 400-μm graphite electrode during the initial cycles at a low rate of C/35.

One of the most recent efforts of the community towards enhanced capabilities of LIBs is focusing on electrode engineering, and specifically examining ways and methods to increase the loadings of the cathode materials and consequently the graphite anode loading and thickness, enabling a higher cell energy. As such, graphite anode thicknesses of 100-600 μm were recently studied and evaluated [19,20,47,48]. The reports indicated the challenges associated with the implementation of ultra-thick electrodes which include diffusion and migration of the Li-ions, tortuosity, electrode architectures, and macro-engineering of their structure. Such issues may hinder complete lithiation of the electrode due to incomplete formation of Li-C phases.

Lithium intercalation in a graphite electrode is generally described by staging [44], where layers between graphene sheets are filled with Li-ions, forming several ordered $Li_xC_6$ phases with 0≤x≤1. These phases can be identified in the recorded diffraction patterns, where in general, even single peaks allow quantifications of phase contents of an electrode. However, Li ions transfer into the formed SEI and low concentration ingress into the crystalline graphite, without distorting the lattice, are missed by diffraction, but are observed by increasing local beam attenuation. Hence, these processes can be identified by correlating transmission and diffraction information, as being enabled by our multi-modal operando approach. The subsequent successive formation and transformation of different $Li_xC_6$ phases can be globally observed by diffraction, whereas correlation with the local attenuation enables quantitative spatially resolved lithiation and delithiation characterization of the formed phases, alongside dead Li and SEI formation observations. These capabilities are crucial to observe the phase formations and associated processes, including SEI formation and dead Li tracing in ultra-thick electrodes to enable a better understanding of the limitation and challenges in operating LIBs utilizing such ultra-thick electrodes.

## 2. Experimental
### 2.1. Materials

A typical LIB half-cell for in-plane neutron imaging and diffraction investigations was designed and a schematic view of the custom-made electrochemical cell is presented in Figure 1. The main features of the cell design are a cylindrical plane layered electrode geometry with interfaces aligned with the neutron beam enabling in-plane imaging and thus depth profile observations with regards to the electrode thickness. An adjustable stack height allows the usage of different electrode thicknesses. The cell casing of Al has an inner diameter of 18 mm, an isolating inner polytetrafluoroethylene (PTFE) layer with a thickness of 1 mm in the active area containing two electrodes with diameters of 16 mm. The thick active graphite electrode had a thickness of 400 µm (projected height, which defines the *y*-direction of the laboratory coordinate system, where *y* = 0 denotes the interface of the graphite electrode with the separator placed between the two electrodes) constituting the region of interest of this study. The second electrode was a metallic Li reference electrode (Li ribbon, 99.9 %, Merck, 0.75 mm thickness) establishing the half-cell. The electrodes were separated by an initially 260 µm thick glass microfiber separator (Whatman Grade GF/A). The three components were immersed in electrolyte (1M lithium hexafluorophosphate in ethylene carbonate:dimethyl carbonate (1M LiPF6 EC/DMC=1:1 vol.%), BASF) and compressed between electrochemically stable current collectors of stainless steel (316SS). The cell was sealed with O-rings, which were fixed between the cell casing, the polyether ether ketone (PEEK) clamping rings and the extended current collectors.

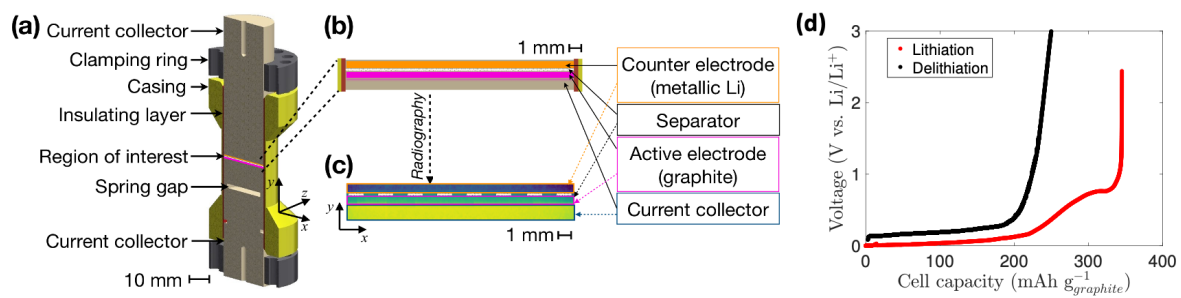

**Figure 1. Experiment** (a) drawing of the custom-designed electrochemical cell; (b) detail of the half-cell configuration as seen by the neutron beam; (c) actual neutron image of the half-cell using a yellow to blue artificial color coding; (d) cycling conditions of the half cell in terms of voltage vs cell capacity.

**2.2. Cycling conditions**

The operando neutron measurement was performed at room temperature. Diffraction patterns and transmission images were recorded for a full discharge/charge cycle of the initially pristine cell. The as-assembled, pristine electrochemical half-cell was operated during the measurement at a constant current in the voltage range of 3.000-0.001–V *vs.* Li/Li$^+$ for one full cycle of the half-cell. This enabled observation of the SEI formation in the graphite electrode from its earliest stages. The voltage profile of the electrochemical cell is plotted in Figure 1(d). The theoretical capacity of the graphite electrode was assumed to be 372 mAh g$^{-1}$, based on LiC$_6$ phase formation, as the fully lithiated state. The cell was cycled at 1.37 mA, corresponding to C/35 cycling rate, as calculated from the theoretical capacity of the graphite electrode.

### 2.3. Neutron diffraction and imaging

Concurrent neutron imaging and diffraction experiments were performed at the SENJU instrument on the beamline BL-18 of the Materials and Life Science Experimental Facility of the Japan Proton Accelerator Research Complex (J-PARC) [49]. SENJU is a dedicated time-of-flight (ToF) single crystal diffractometer by design providing a pulsed polychromatic neutron beam with a selected wavelength range from 2.0-6.4 Å of a Maxwellian spectrum peaking at 2.3 Å for our combined near-field transmission imaging and far field diffraction study. To this end the instrument was retrofitted with an imaging detector complementing the use of the 90 deg detector modules of the standard set-up of SENJU. The diffraction detector modules are placed 0.8 m from the sample position and feature a pixel size of 4x4 mm$^2$. Together with the ToF wavelength resolution d$\lambda$/$\lambda$<1% they provide a crystal lattice d-spacing resolution of $\Delta$d/d < 10$^{-2}$. Each of the two symmetric 90 deg detectors consists of three 64x64 pixel scintillator area detectors stacked vertically.

Neutron ToF diffraction enables efficient crystallographic measurements with a polychromatic beam and a position sensitive area detector. The wavelength resolution is given by the time-of-flight, the angular resolution by the divergence, and the detector geometry. The diffraction signal can be related to d-spacing through the Bragg equation d$_{hkl}$=n$\lambda$/(2sin$\theta$), where $\lambda$ is the wavelength, $\theta$ the diffraction angle, and hkl are the Miller indices of a specific family of lattice planes. Based on the chosen spectral range, the Bragg diffraction detected in this geometry covers a d-spacing range of 1.7-4.4 Å; thus, the most intense diffraction peaks of graphite and its ordered lithiation phases Li$_x$C$_6$, i.e. C(002), LiC12(002) and LiC6(001) ranging between 3.35 and 3.71 Å. Li is intercalated in graphite in stages of filling layers between the graphene sheets forming distinct ordered Li$_x$C$_6$ (0≤x≤1) lithiation phases which can be depicted by diffraction due to their specific lattice parameters. Individual peaks are sufficient to distinguish these phases. Fig. 2(a) displays the diffraction patterns measured throughout cycling. The intercalation stages can be defined and identified consecutively as follows [50]: stage IL when Li ions are initially intercalated in a not yet ordered but rather liquid-like manner with $0<x\leq0.2$ ($LiC_{30}$); stage IV is corresponding to $0.2<x\leq0.25$ ($LiC_{24}$); stage III with $0.25<x\leq0.33$ ($LiC_{18}$); stage IIL and stage II with $0.33<x\leq0.5$ ($LiC_{12}$); and finally stage I forming $LiC_6$ with $0.5<x\leq1$. The initial stage IL is overlapping in d-spacing with graphite but is characterized by an eventually growing d-spacing. Similarly, the so-called intermediate stages IV/III and IIL overlap amongst each other and are difficult to distinguish clearly by diffraction solely despite increasing lattice spacings. $LiC_{12}$ and $LiC_6$ on the other hand provide clearly distinct diffraction peaks (Fig. 2) and their individual phase content can be deduced straightforwardly from the peak intensities, based on the known crystallography [51,52] when accounting also for attenuation, which increases significantly with Li content.

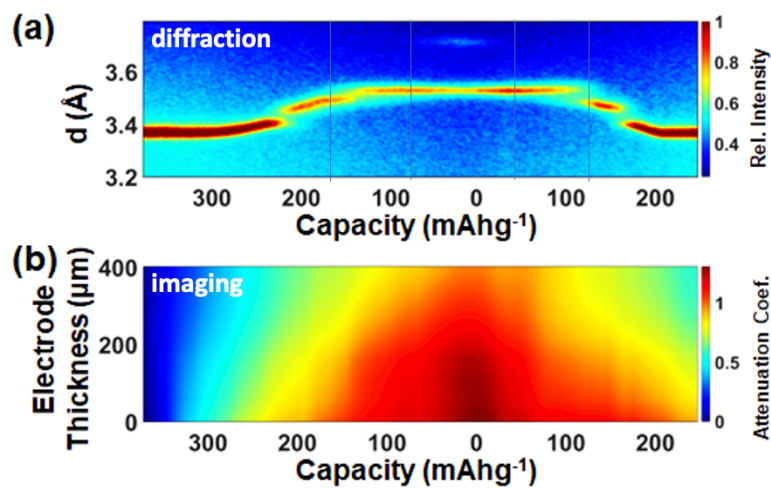

**Fig. 2 Reduced data** measured in concurrent neutron diffraction and imaging experiment; (a) diffraction patterns recorded throughout lithiation and delithiation; (b) horizontally averaged change of attenuation coefficient versus electrode thickness (distance from separator) throughout lithiation and delithiation;

For transmission imaging a micro-channel plate (MCP)/TPX type imaging detector was placed in the near field right behind the sample cell [45]. It has a field of view of $28\times28mm^2$ divided into 512x512 pixels with a 55 μm pixel pitch. The neutron beam cross section was limited to 18 x 2.4 mm for full field illumination of the half-cell, i.e. the full half-cell width as well as the full height of the active area containing the electrodes the separator and part of the current collector. The near field imaging detector was also utilized to carefully align the cell, i.e. the plane of the Li electrode with the beam. The spatial resolution of the recorded images is defined by the imaging geometry, i.e. the beam divergence and the distance between sample and detector [53] which was of the order of a few millimeters. Together these imply an achievable image resolution of about 100 μm in this given case.

The contrast of the images recorded changes depending on the Li redistribution in the cell, with Li ions having a particularly high neutron attenuation cross section (~70 barn). The wavelength dependent attenuation for different lithiation phases of graphite was calculated utilizing the program ncrystal [54]. The development of the attenuation for different lithiation states of graphite within the utilized wavelength range displays an approximately linear scaling with Li content, which dominates the attenuation. The evolution of the local attenuation coefficient μ(x,y,t) depicted in imaging with time can be described by the Beer-Lambert law as:

$$\mu(x,y,t) = -\frac{1}{D(x)} \ln\left(\frac{I(x,y,t)}{I_0(x,y,t)}\right).$$

The transmission thickness D(z) is calculated straightforwardly according to the known cylindrical shape of the cell. It has to be noted that the calculated attenuation coefficient μ(x,y) here would not be suited to characterize a specific material, as it integrates all material phases along the beam path, including the Al casing and the electrolyte which is dispersed throughout the cell. For analyzing the process in terms of Li distribution, thus, the initial image of the pristine cell at t=0 has been subtracted from all subsequent images. Hence, only changing contributions remain which are assumed to equal the uptake and release of Li, when the focus is solely on the graphite layer. Despite the images allowing to assess this process locally across the electrode, the observations presented here will be limited to the integral of a wide central regime of the width of graphite electrode. The focus is on observing the evolution μ(y,t) during cycling with the spatial resolution concerning solely the distance y from the separator. To this end Fig. 2b illustrates the evolution of the attenuation during cycling as a function of distance from the separator versus capacity. For these representations the attenuation coefficients have been averaged across the respective width of the electrode. This is not only for visualization purposes, but also related to the focus of the study on the evolution across the thickness of the electrode.

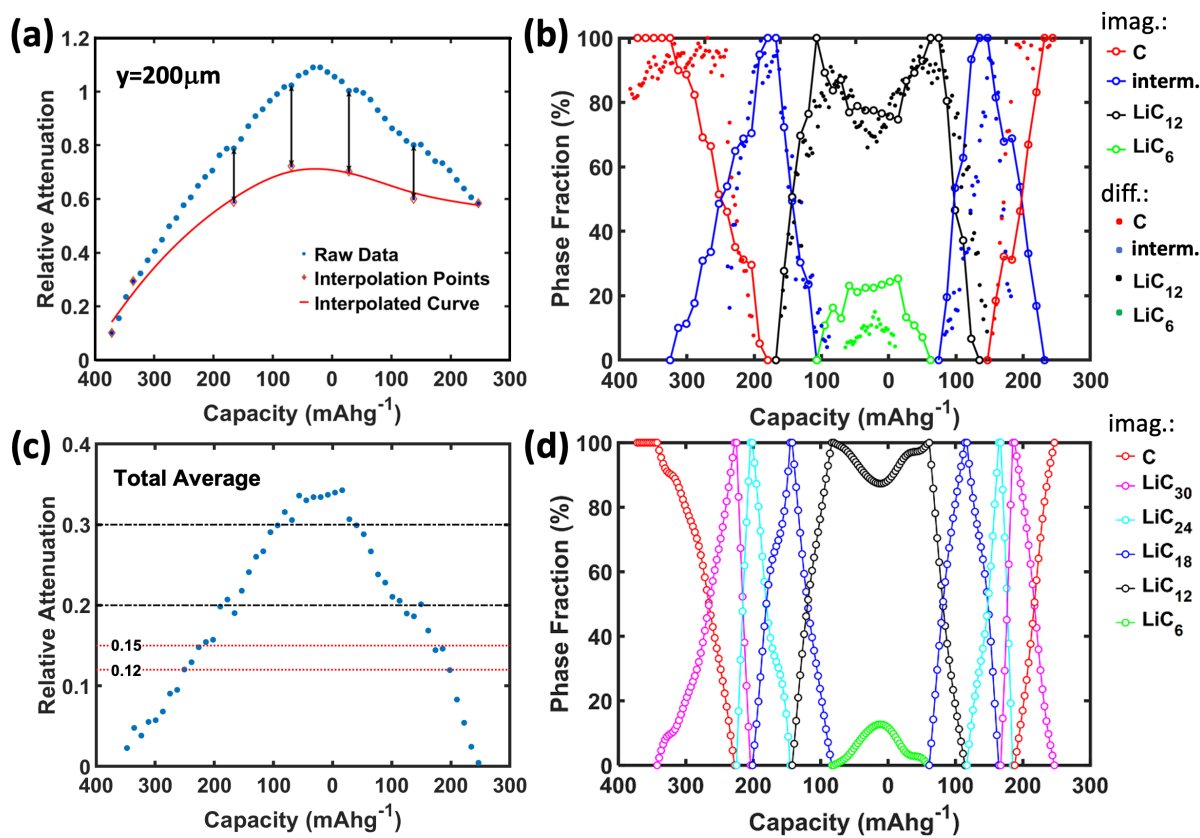

**Fig. 3 Correlative data analysis**; (a) profile of the attenuation evolution in the graphite electrode throughout cycling indicating the points of phase saturation for the intermediate phases and LiC$_{12}$, as extracted from diffraction; the red curve is the interpolation of contributions other than from the intercalation of Li; (b) evolution of phase fractions throughout cycling from diffraction and the analyses of imaging based on (a); (c) profile of attenuation of the electrode throughout cycling after subtraction of the interpolated contributions of non-intercalated Li; the horizontal lines indicate the attenuation values indicative of saturated lithiation phases; (d) evolution of phase fractions extracted from imaging when accounting also for the individual intermediate phases;

## 2.4 Data analysis

The phase fractions of the ordered lithiation stages are deduced from analyses of the recorded diffraction peaks and are displayed as colored dots in Fig. 3 in the top right panel. The total time dependent transmission of the graphite electrode was extracted from the integral of the corresponding area in the near field transmission data and was used to normalize the phase fractions extracted from the diffraction peaks. Accordingly, the final evaluated phase fractions add to about unity throughout the measurement and reveal the integral lithiation behavior of the electrode during cycling as reported previously [55,56]. As only ordered Li$_x$C$_6$ phases contribute to the diffraction signal, diffraction alone is blind to Li ions in the electrolyte and/or potential disordered solid compounds in or on the solid graphite matrix of the electrode, such as in the SEI. However, our concurrent imaging approach enables to additionally observe all other Li fractions such as dead Li and reversibly plated Li. Further analyses of the transmission data assume that overall, the redistribution of Li ions is solely due to the following processes: lithiation/delithiation process, Li plating, and the formation of the SEI. By that, one should realize that dead, plated and active lithium, are responsible for significant changes in the attenuation coefficient.

## 3. Results and Discussion

A time dependent line profile $\mu(y,t)$ at y=200 µm is presented in Figure 3a, illustrating how the attenuation of non-intercalation phases, such as the SEI, can be estimated based on the points of individual saturated lithiation phases, known from diffraction and the known attenuation of such state. The diffraction data reveals at which points in time in the process the ordered intermediate stage III (LiC$_{18}$) and stage II (LiC$_{12}$) reach completion throughout the electrode bulk, while no concurrent phase is present. Subtraction of interpolated curves based on these correlations result in attenuation profiles, solely originating from Li ions intercalation into the graphite electrode, as shown in Fig. 3(c). Subsequently, the attenuation coefficients at each time and layer of the electrode can be evaluated in terms of lithiation phase fractions, corresponding to the phase fraction analyses of the diffraction data. A comparison in Fig. 3(b) shows a good agreement of both results, when taking the full examined graphite electrode region into account. Fig. 3(b) shows deviations e.g. at the disappearance of the pure graphite phase between diffraction and transmission results. These deviations can be attributed to the lacking sensitivity of diffraction to the non-ordered lithium intercalation at this stage because in contrast to imaging diffraction only probes lithium intercalation that affects the lattice spacing. Thus, the contribution of the concurrent method of transmission together with diffraction enables sensitivity to the kinetics of Li intercalation even at higher stages than LiC$_{18}$. On the other hand, deviations in the LiC$_6$ phase fraction at increased Li-ion intercalation regions might be due to an underestimation of the faint LiC$_6$ phase diffraction peak at correspondingly high attenuation of the increased Li ion content in the electrode. The attenuation profiles enable consideration of additional intermediate phases, which are not distinct in the diffraction. Thus, a more detailed phase evolution including transient $LiC_{30}$, $LiC_{24}$ and $LiC_{18}$ phases, can be depicted, as shown in Fig. 3d. Fig 3d also allows determination of the specific capacities in which the full different Li phases are reached. In this context it is also noteworthy that transient plateaus in the attenuation profiles indicate a saturation of individual lithiation stages (Fig. 3c), resembling similar effects, reported in the voltage profiles of the (de-)intercalation process.

The fact that these phase evolutions cannot only be depicted integrally (like in diffraction and in Fig. 3), but can be resolved locally, enables mapping the lithiation phases as a function of the distance into the electrode, from the separator to the current collector, and as a function of cycling time (being related to the overall charge capacity). Such phase maps for the individual lithiation phases are presented in Figure 4. These maps show the kinetics of lithiation and delithiation processes not only with overall capacity changes but also with spatial variations relative to the electrode thickness. The indication is of a relatively homogeneous lithiation and delithiation across the thick electrode occurring at relatively low C-rate. The initial Li-ion uptake in the graphite electrode shows a slight directional signature, of increasing Li-ion concentrations, evolving from the separator side. This gradient changes with subsequent lithiation phases, appearing slightly earlier at the vicinity of both electrode surfaces, until the $LiC_{12}$ phase shows an onset at the separator side, while the final $LiC_6$ phase develops a relatively homogeneous region across the whole electrode thickness, with a faint maximum in the central region. Similar smooth trends can be observed during delithiation, though a trend of slower phase transformation to the subsequent phase, close to the surface regions of the electrode is observed for $LiC_{12}$ and $LiC_{18}$.

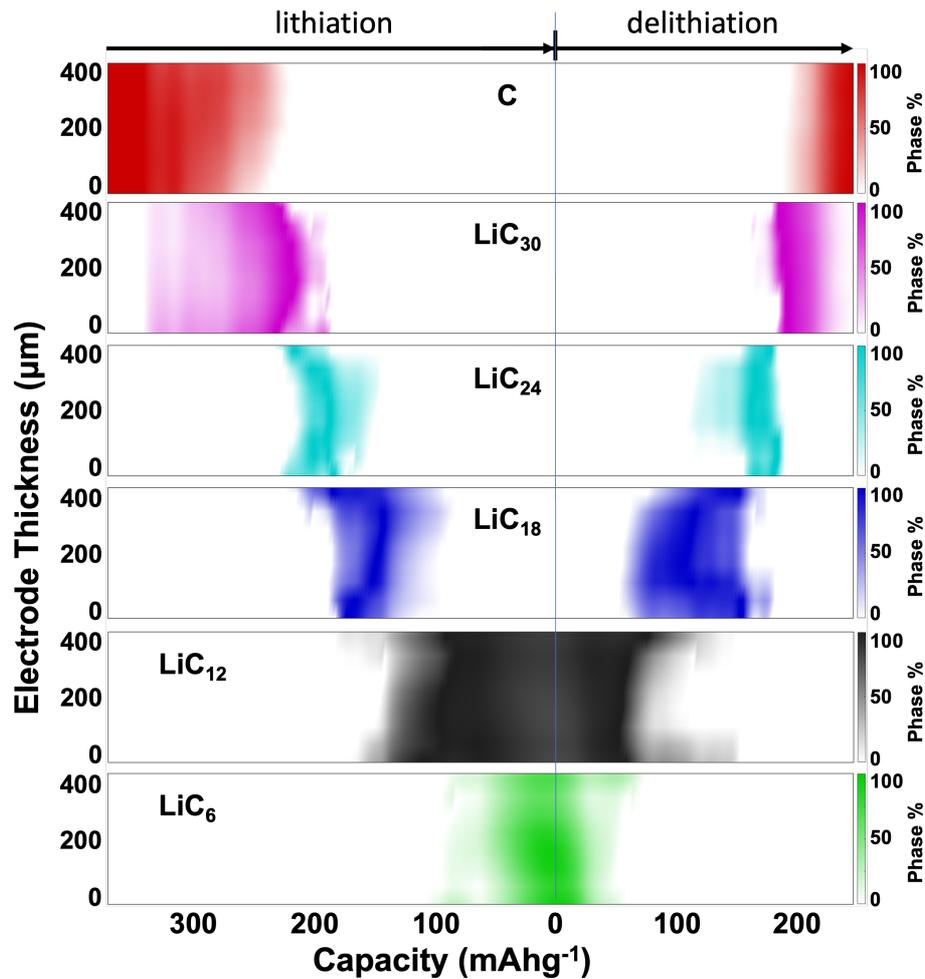

**Fig. 4 Phase evolution against electrode thickness throughout cycling**; the individual panels illustrate the evolution of an individual lithiation phase each during cycling of the battery and spatially resolved for the thickness of the electrode;

Turning the attention to the interpolated contributions to the image contrast during cycling presented in Fig. 5 reveals both, clear reversible and irreversible contributions. While Fig. 2b presents all contributions of Li to the evolving attenuation, Fig. 5 displays specifically the contributions not intercalated Li, e.g. dead lithium, SEI, and potentially also plated Li metal. The similarity of Fig. 5 with Fig. 2b is owed to the strong contribution of these fractions of Li and their distinct distribution as compared to a relatively homogeneous distribution of the intercalated lithium with much more faint variations which are depicted in Fig. 4. The reversible contribution of the non-intercalated fraction is found to be of the order of 20% compared to the contribution of attenuation due to Li-ion intercalation. The irreversible contribution can be associated with the formation of the SEI and dead lithium, while the reversible contribution is assumed to represent plated lithium and a potentially reversible SEI phase. The main fraction of non-intercalated Li is being concentrated in the vicinity of the separator/electrode interface with a distinct gradient towards the Cu current collector. Such spatial trend is observed for reversible and irreversible contributions. Not distinguishing these contributions from the intercalation process would imply a misleading interpretation of the recorded images, which would wrongly imply that the Li ion intercalation process is significantly more pronounced at the separator side of the electrode.

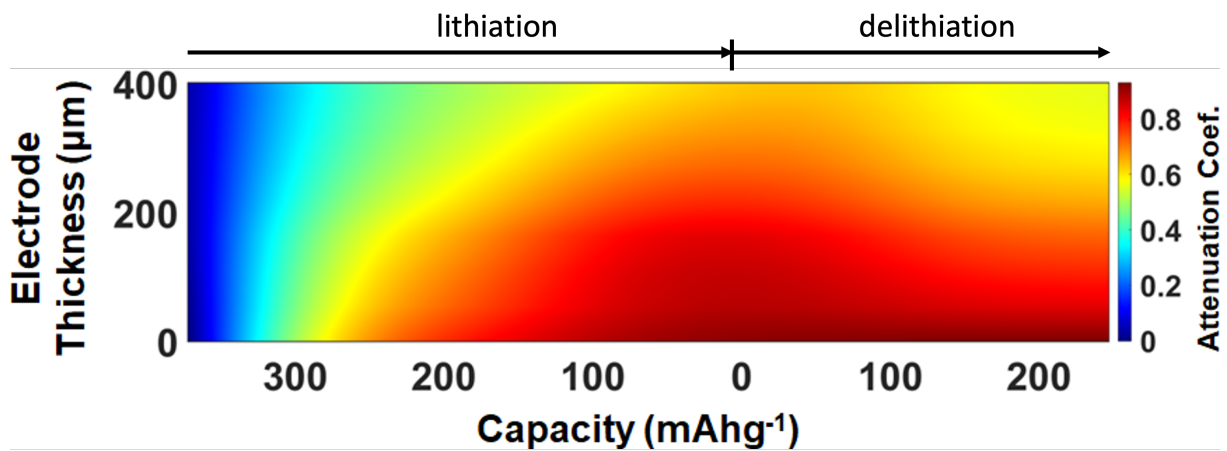

**Fig. 5 Evolution of non-intercalation related attenuation contrast;** the vertical axis is the electrode thickness by distance from the separator, while the horizontal axis indicates the capacity during lithiation on the left of zero and delithiation on the right side;

The residual irreversible contribution, on the other hand, is likely to contain lithiation phases of $LiC_{>30}$ which are described by Takagi et al. [57] and cannot be identified by the presented measurement techniques. Their presence is, however, supported when considering the large fraction of irreversible capacity (Fig. 6), which amounts to about 100 mAh g$^{-1}$, of which only about 40 mAh g$^{-1}$ can be associated to the SEI formation in the potential window of 0.9-0.7 V vs. Li/Li$^+$. Such fraction of irreversible capacity in a range of 5-15 % of the total capacity is typically associated with SEI build-up on common graphite composite electrodes in the alkyl-carbonate based electrolyte [58]. The remaining 60 mAh g$^{-1}$ irreversible capacity coincides with values provided being associated with residual low lithiation phases of $LiC_{36}$ and $LiC_{72}$ [57]. We therefore conclude that the excess of accumulated irreversible capacity upon switching the electrode polarization from Li intercalation to de-intercalation, is because a fraction of Li-ions corresponding to amounts and capacity of the postulated $LiC_{36}$ and $LiC_{72}$ phases are being trapped and consequently being immobilized. A similar phenomenon was reported for fast charging of graphite anodes [23] where upon charging with a rate of 6C the graphite anode exhibited a significant heterogeneity, due to the incomplete and inhomogeneous intercalation reactions hindered by rather sluggish reaction kinetics. It was concluded that Li ion interface diffusion dominates the reaction kinetics at high rates in thin graphite electrodes. As Li ions diffusion challenges cannot be neglected for a thick graphite electrode, these would suggest corresponding phenomena to cause irreversible capacity losses. Here, we have utilized an ultra-thick electrode of 400 μm cycled at a very low C rate (C/35), and one may, thus, conclude that the obstacles and challenges faced by the utilization of ultra-thick electrodes correspond to those related to fast-charging electrodes. Thus, also in this case, means to overcome such shortcomings might be found in special electrode designs and architectures, including e.g. advanced 3D current collectors.

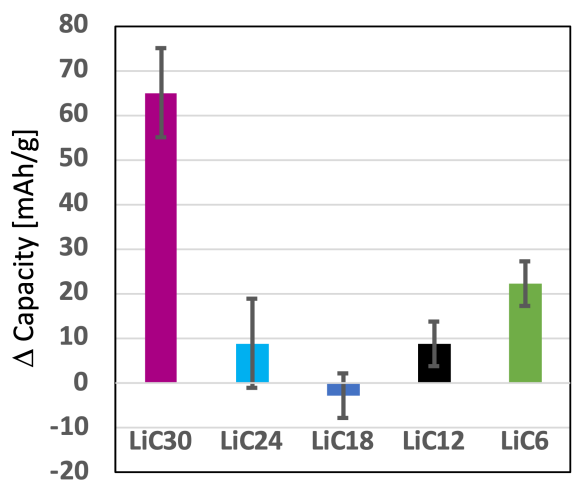

**Fig. 6 Differential capacity** ($\Delta$Capacity) with regards to forming and vanishing lithiation phases

## 4. Conclusions

Through this work we provide a powerful advanced operando characterization tool for spatially resolved observations of the evolution of lithiation stages through full field imaging of ultra-thick and advanced Li-ion graphite anode battery electrodes. Diffraction guided neutron imaging enables distinction between different contributions by correlating the detected image contrast to the diffraction information in *an operando* measurement. Namely, the intercalation process of Li-ions can be distinguished and separated from the SEI formation process, and dead lithium. This allows for the study of the lithiation and delithiation phases of the electrode with spatial resolution, revealing their time and depth dependent evolution. To this end, we find that different lithiation stages co-exist during charging and discharging of an ultra-thick graphitic-based electrode. We observed a delayed lithiation and delithiation processes at the central region of the electrode, while SEI formation, and dead lithium are predominantly found closer to the interface with the separator. In addition, the results allow us to postulate that irreversible capacity is associated with trapped Li ions in low concentration Li-C phases. The study furthermore underlines the potential of the method to study Li diffusion and lithiation phase kinetics in advanced thick electrodes, and we expect that the method can be extended also to investigations of fast charging with increased spatial and temporal resolution at advanced neutron imaging and diffraction beamlines.

**Methods**

Electrode preparation

The ultra thick composite graphite-based electrode was prepared in three steps. First, 88 wt.% natural graphite powder (Formula BT, SLC 1520P, Carbo Tech Nordic ApS) as the active material was dry mixed in a mortar with 3 wt.% of conductive carbon (carbon black, Cabot Corporation, USA), and 9 wt.% of PTFE (Polytetrafluoroethylene, Sigma-Aldrich) as the binder. Second, the homogeneous powder mixture was pressed into a 16 mm-diameter disk pellet by applying 3 tons of pressure with a SPECAC press. Finally, the pellet was dried in a BÜCHI Glass Oven B-585 vacuum furnace within a glovebox filled with Ar atmosphere. The temperature in the oven was set to 80◦C for 4 hours and was further increased to 120◦C for 10 hours. The pellet was kept under Ar atmosphere thereafter until the electrochemical cell was assembled, i.e. a day prior to the experiment, allowing electrolyte to disperse through the porous electrode.

Neutron Measurements

The neutron flux was $3 \times 10^7$ n s$^{-1}$cm$^{-2}$ at the sample position of SENJU at 34.8 m from the source operated at a power of 500 kW [49]. The pulsed neutron source allows to obtain crystallographic phase information by time-of-flight (ToF) diffraction probing a sufficient instantaneous scattering vector range on the θ = 90+/-9 degree detector banks to observe the respective lattice spacings d=λ/(2sinθ) of the involved carbon lithiation phases integrally throughout the electrode. At the same time the transmitted polychromatic ('white') beam provides information on the local Li concentration in the electrode resolved by an MCP/timepix imaging detector [59] with 55 micrometer pixel size. The diffraction data was recorded continuously in a time stamped list mode data acquisition, whereas transmission images were recorded with 15 min time resolution. Subsequently the diffraction data was reduced to 15 min time resolution to match the applied imaging time resolution. The diffraction data was processed and analysed with the software STARGazer [60]. The operando cycling of the half-cell in the experiment was performed using a CompactStat electrochemical workstation by Ivium.

**Acknowledgments**

M.E.B, R.E.J, S.S. A.T., and L.T.K. thank the Danish Agency for Science, Technology and Innovation for funding the instrument centre DanScatt supporting beamtime travels. M.S. acknowledges the financial support from Otto Mønsted's Fund. P.K.M.T. thanks for financial support by OP RDE, MEYS, under the project "European Spallation Source - participation of the Czech Republic - OP," Reg. No. CZ.02.1.01/0.0/0.0/16_013/0001794. I.G. and Y.E.E. thank the financial assistance and support provided by the PAZY Foundation, under Grant contract # 5100068039. Y.E.E. thanks the infrastructure support provided by the Israel National Institute for Energy Storage (INIES) and Grand Technion energy Program (GTEP). We acknowledge beamtime provided by JPARC to perform the measurements at SENJU under the Proposal number 2018B0076 and 2018L0501. The authors thank Dr. Takashi Ohhara, Anton S. Tremsin, Takenao Shinohara for experimental support. The authors thank Rune E. Johnsen, Christian Baur and Sigita Trabesinger for valuable scientific discussions.